\documentclass{PoS}

\usepackage[font=small,labelfont=bf]{caption}

\title{Multi-Messenger Connections among High-Energy Cosmic Particles}

\ShortTitle{Multi-Messenger Connections among High-Energy Cosmic Particles}

\author{\speaker{Kohta Murase}\thanks{}\\
        Department of Physics; Department of Astronomy \& Astrophysics; Center for Particle and Gravitational Astrophysics, 
        The Pennsylvania State University\\
        Yukawa Institute for Theoretical Physics, Kyoto University\\
        E-mail: \email{murase@psu.edu}}

\abstract{
The origin of high-energy cosmic neutrinos is one of the biggest mysteries in astroparticle physics. The fact that diffuse intensities of high-energy neutrinos, ultrahigh-energy cosmic rays, and GeV-TeV gamma rays are all comparable suggests that these messengers are physically connected. The IceCube data above 100 TeV energies can be naturally explained by cosmic-ray reservoir models. In particular, starburst galaxies and galaxy clusters/groups serve as natural storage rooms of cosmic rays, and it has been theoretically predicted that these sources are promising sites of high-energy neutrinos and gamma rays that are produced via inelastic $pp$ interactions. Indeed, the predictions made before the discovery of IceCube neutrinos are consistent with the current high-energy neutrino data measured in IceCube, and that they could give a grand-unified view of sub-PeV neutrinos, sub-TeV gamma rays, and ultrahigh-energy cosmic rays. These unified models have strong prediction powers, which can be tested by next-generation neutrino detectors such as IceCube-Gen2 as well as gamma-ray telescopes such as CTA. 
The recent observations have also shown that the 10-100~TeV diffuse neutrino flux is higher than that at PeV energies, which suggests that they come from a different class of neutrino sources. The detailed comparison with the diffuse isotropic gamma-ray background measured by {\it Fermi} has revealed that these medium-energy neutrinos are likely to come from hidden cosmic-ray accelerators, from which neutrinos can escape while GeV-TeV gamma rays are attenuated. The candidate source classes are choked gamma-ray burst jets and active galactic nuclei (AGN) cores. In particular, the AGN corona model predicts a unique connection between 10-100 TeV neutrinos and MeV gamma rays, which can be robustly tested with future MeV gamma-ray missions such as AMEGO.   
}

\FullConference{36th International Cosmic Ray Conference -ICRC2019-\\
		July 24th - August 1st, 2019\\
		Madison, WI, U.S.A.}

\begin{document}

\section{Introduction}
The origin of ultrahigh-energy cosmic rays (UHECRs) has been a long-standing mystery. Candidate sources are extraordinary objects like gamma-ray bursts (GRBs) and active galactic nuclei (AGN), which also involve many unresolved problems in source physics. The source identification should largely help us understand their source physics. UHECR observations provide important information of their origins, which include: (i) the energy flux spectrum, (ii) anisotropy of arrival directions, and (iii) nuclear mass composition. The observed cutoff in the UHECR energy spectrum suggests an extragalactic origin, but it can be explained by various possibilities. The arrival direction of UHECRs is potentially useful because UHECRs experience only small deflections as a consequence of their extreme energies. However, the attempts have not been successful so far.  

The breakthroughs occurred recently.  In astrophysical sources, neutrinos are produced by cosmic rays via inelastic $pp$ collisions with surrounding gas or $p\gamma$ interactions with ambient photons. They provide a unique probe of cosmic-ray accelerators, since they can reach the Earth without deflections by cosmic magnetic fields or attenuation due to interactions. The IceCube Collaboration reported the discovery of cosmic high-energy neutrinos in the PeV range~\cite{Aartsen:2013bka}.  The follow-up analyses established that the observed events should not be entirely backgrounds from interactions between cosmic rays and the Earth's atmosphere.  The diffuse neutrino intensity for the sum of all three active neutrino flavors is about $3\times10^{-8}~{\rm GeV}~{\rm cm}^{-2}~{\rm s}^{-1}~{\rm sr}^{-1}$ in the sub-PeV range~\cite{Aartsen:2013jdh}.  

The detection of cosmic high-energy neutrinos opened up a new window on the Universe, but it also brought new puzzles. So far, their arrival directions are consistent with an isotropic distribution, suggesting an extragalactic origin. The present data show no evidence for event clustering around known astrophysical objects (except for two tentative candidates, TXS 0506+056 and NGC 1068), requiring source classes that are more abundant than blazars. The origin of high-energy neutrinos has emerged as a new mystery in particle astrophysics~\cite{Murase:2016gly}. 

Probably one of the most important facts is that the observed energy flux of neutrinos in the $0.1-1$~PeV range turned out to be comparable to those of the diffuse sub-TeV gamma-ray flux and the UHECR flux (see Fig.~1 left). This fact implies that the energy budgets of three messenger particles are somehow all comparable, even though their energies range over ten orders of magnitude~\cite{Murase:2018utn}. It is natural to ask whether this coincidence is physical or accidental. The grand-unified scenario, in which all three messenger particle fluxes are explained by a single population of the sources, has been proposed~\cite{Murase:2016gly,Fang:2017zjf}, and starburst galaxies and galaxy clusters/groups have been considered as primary candidate sources~\cite{Loeb:2006tw,Thompson:2006np,Murase:2008yt,Kotera:2009ms}. 

The latest IceCube data have also revealed another puzzling feature. Shower analyses allow us to study the neutrino flux below 0.1 PeV, which showed that the $10-100$~TeV neutrino flux is as large as $\sim10^{-7}~{\rm GeV}~{\rm cm}^{-2}~{\rm s}^{-1}~{\rm sr}^{-1}$, with a steeper spectral index~\cite{Aartsen:2014muf,Aartsen:2015ita,Aartsen:2017mau}. 
This flux level is significantly higher than that for neutrinos above 100~TeV energies. The detailed comparison with the diffuse isotropic gamma-ray background indicates the necessity of hidden neutrino sources from which gamma rays cannot escape (see Fig.~1 right). This is because gamma rays associated with neutrinos overshoot the {\it Fermi} data if the sources are optically thin to the two photon annihilation process.  However, there are not many possibilities that can explain the data, deepening the mystery on their origin. The possible candidate sources AGN cores and choked GRB jets~\cite{Murase:2015xka}. 

It is crucial to identify the sources of high-energy neutrinos and understand connections among different cosmic messenger particles. In this paper, we discuss cosmic-ray reservoir models that may be responsible for $\gtrsim100$~TeV neutrinos as well as the sub-TeV gamma rays and UHECRs, and hidden neutrino sources that are required to explain the $10-100$~TeV neutrino data.

\begin{figure}[tb!]
\begin{center}
\begin{tabular}{cc}
\includegraphics[height=5.0cm]{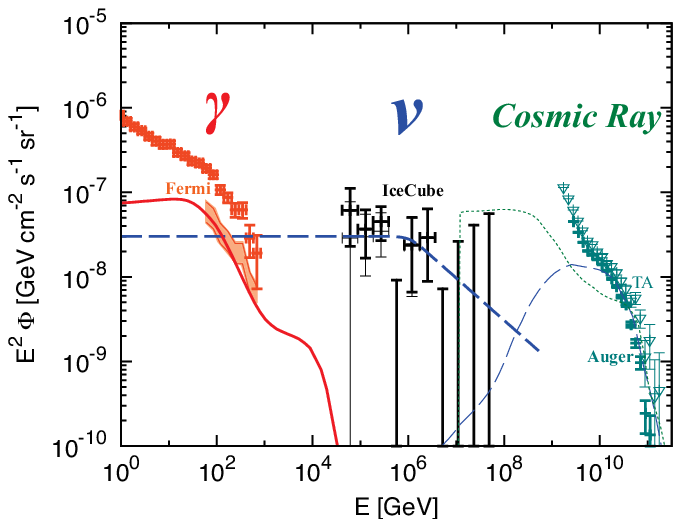} 
\includegraphics[height=5.0cm]{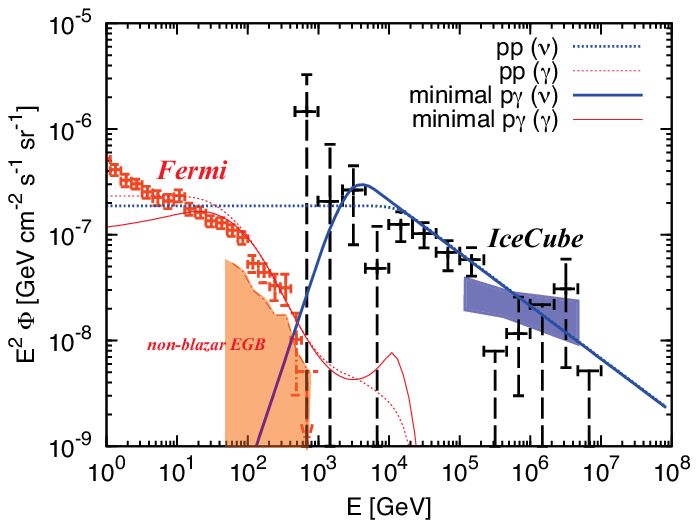} 
\end{tabular}
\end{center}
\vspace{-1.5em}
\caption{
\textsl{Left: High-energy diffuse fluxes of three messenger particles~\cite{Murase:2016gly}. One sees that the energy fluxes of sub-TeV gamma rays, PeV neutrinos, and UHECRs are all comparable, while particle energy spans over ten orders of magnitude. These observations can naturally be accommodated by cosmic-ray reservoir models, in which UHECR accelerators with a hard spectral index of $s\sim2$ are embedded in gaseous environments such as starbursts and galaxy clusters/groups.\\
Right: The diffuse neutrino flux measured by upgoing muon neutrino searches (blue shaded) and shower analyses (black data points), as well as the diffuse gamma-ray flux (red data points) and its non-blazar contribution (red shaded). The medium-energy neutrino data violate the non-blazar component of the extragalactic gamma-ray background if the sources are optically thin to the two-photon annihilation process~\cite{Murase:2015xka}.}}
\vspace{-0.5em}
\label{fig:cosmicray}
\end{figure}

\begin{figure}[tb!]
\begin{center}
\begin{tabular}{cc}
\includegraphics[height=5.0cm]{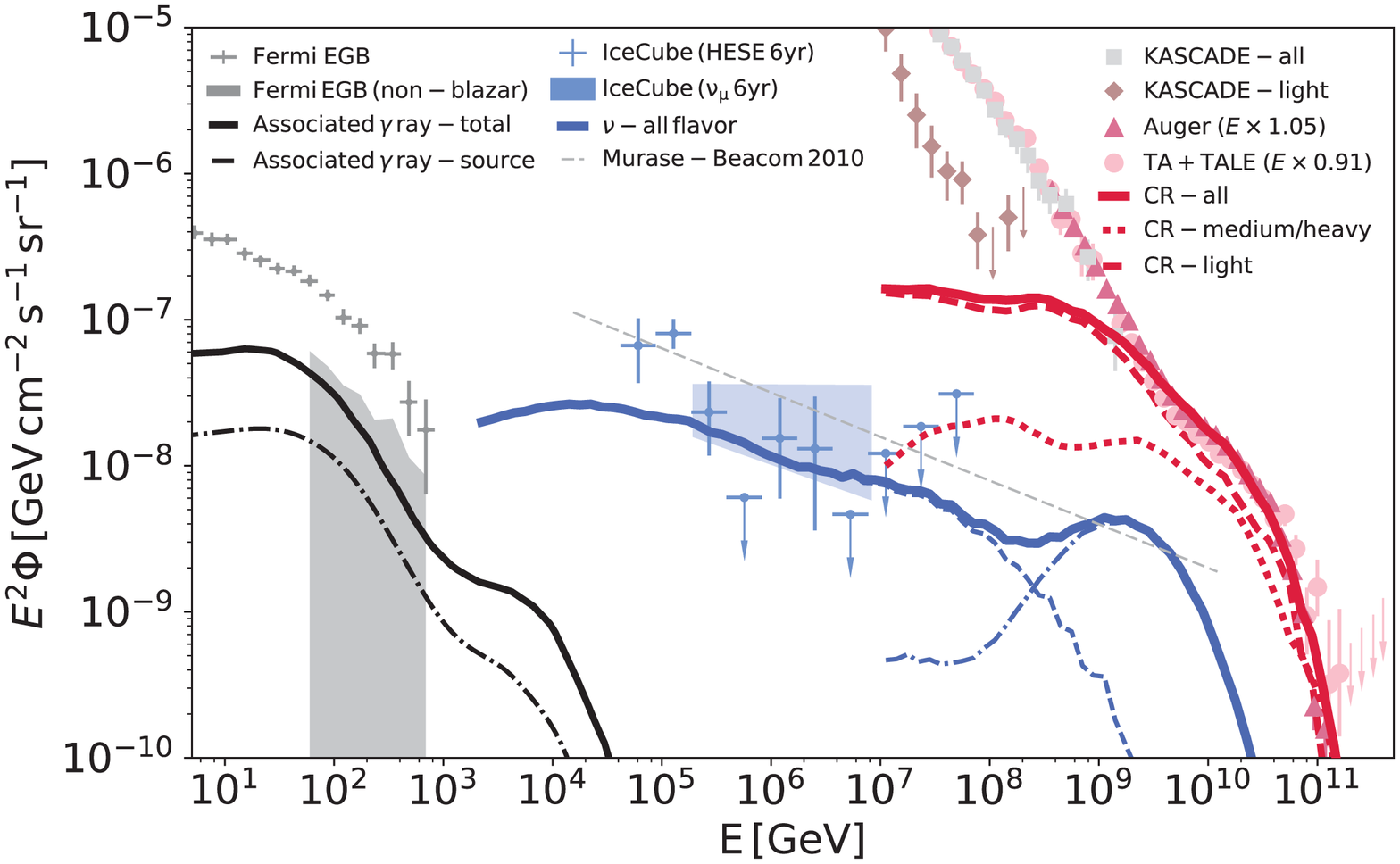} 
\includegraphics[height=5.0cm]{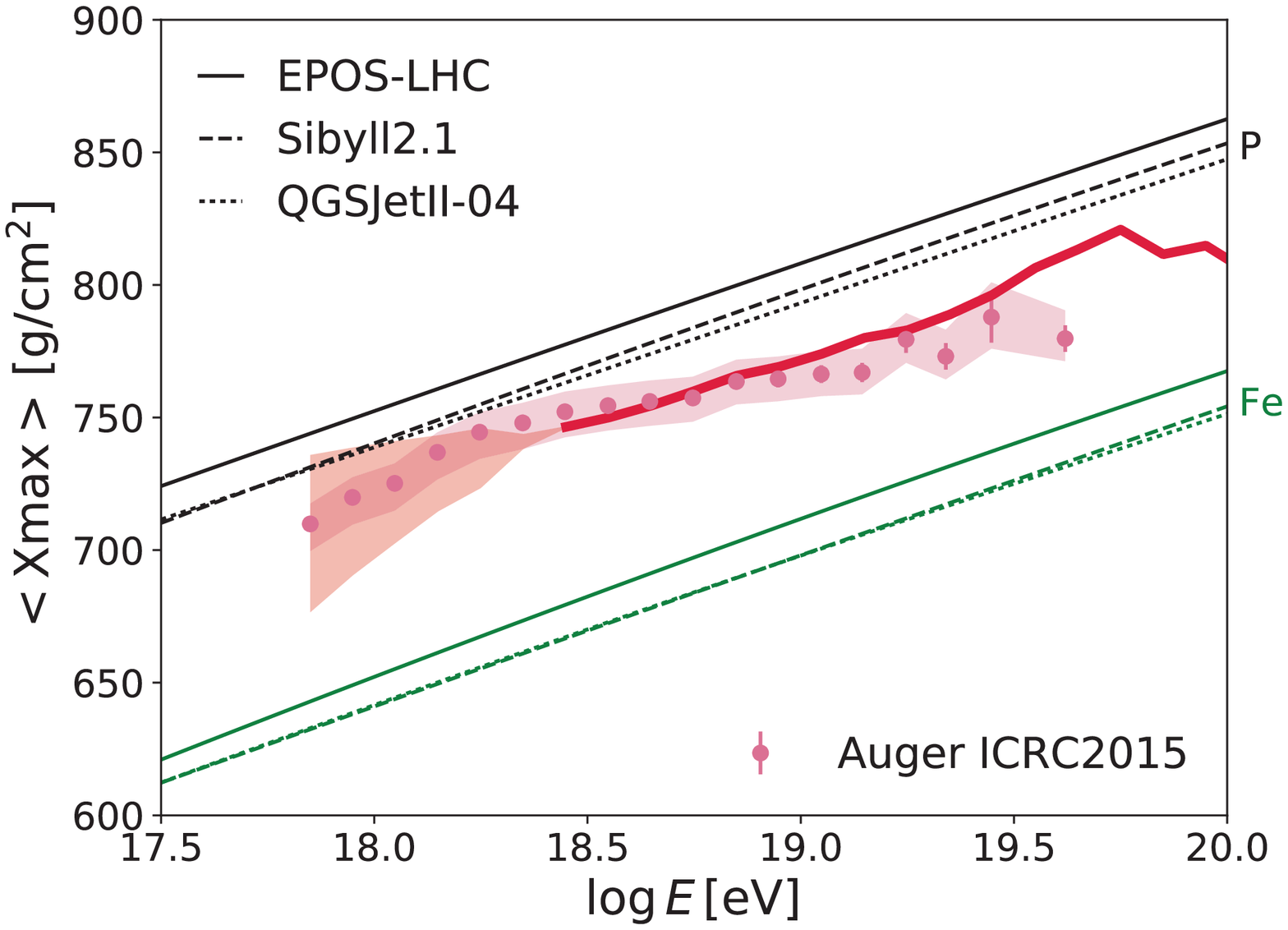} 
\end{tabular}
\end{center}
\vspace{-1.5em}
\caption{
\textsl{Left: Predictions of the galaxy cluster/group model for sub-PeV neutrinos, sub-PeV gamma rays, and UHECRs~\cite{Fang:2017zjf}. The total cosmic-ray spectrum (solid red), the spectrum of light nuclei (dashed red; H and He) and that of medium/heavy nuclei (dotted red; CNO, Si, Fe) are shown. Neutrinos (solid blue) are produced by interactions between cosmic rays and the intercluster material (dashed blue) and those between UHECRs and the cosmic background radiation during their intergalactic propagation (dash-dotted blue). 
The nucleus-survival bound for a cosmic-ray spectral index with 2.3 is also shown (dashed grey). 
The gamma-ray fluxes (solid black for the total contribution and dash-dotted black for the intracluster contribution) are comparable to the non-blazar component of the extragalactic gamma-ray background.\\
Right: Mean of the maximum depth of an air shower of UHECRs~\cite{Fang:2017zjf}. The red shaded region indicates the energy range where the extragalactic contribution is less than 85\% of the measured flux, where the residual component is attributed to Galactic sources.
}
}
\vspace{-0.5em}
\label{fig:cosmicray}
\end{figure}

\section{Cosmic-Ray Reservoirs and Grand-Unification of High-Energy Cosmic Particles}
The fact that energy fluxes of high-energy neutrinos, gamma rays, and UHECRs are all comparable suggests there may be a physical link among three messengers. 
Refs.~\cite{Murase:2016gly,Fang:2017zjf} proposed a grand-unified picture of these three messengers, in which the diffuse fluxes of these particles originate from the same class of astrophysical sources. In general, there are two classes of sources: cosmic-ray reservoirs and cosmic-ray accelerators. In the former class of models, cosmic-ray accelerators are embedded in environments, and neutrinos and gamma rays are produced in environments by cosmic rays escaping from the accelerators~\cite{Murase:2013rfa}. The candidate sources include galaxy clusters/groups~\cite{Murase:2008yt,Kotera:2009ms} and starburst galaxies~\cite{Loeb:2006tw,Thompson:2006np,Tamborra:2014xia}. In these reservoir models, high-energy neutrinos produced by confined cosmic rays are responsible for the IceCube data, while sufficiently high-energy cosmic rays diffusively escape into the intergalactic space without efficient hadronic interaction losses. If the accelerators are UHECR emitters, the observed UHECR flux may be explained. Gamma rays are produced both inside the environments and in intergalactic space (as cosmogenic gamma rays), which contribute to the non-blazar component of the diffuse gamma-ray flux via electromagnetic cascades. 

Radio-loud (RL) AGNs, including blazars and radio galaxies, have been considered as the most promising site of UHECR acceleration. It is not trivial to explain the observed heavy composition in typical AGN models, but it has recently proposed that galactic cosmic-ray nuclei can be reaccelerated to ultrahigh energies by shocks and/or shear in the jet-cocoon system. In addition, UHECRs might be produced by cluster accretion shocks and merger shocks. 
Lower-energy cosmic rays should confined in radio bubbles of radio galaxies, whereas high-energy cosmic rays escaping AGNs should still be trapped in galaxy cluster environments for cosmological time. As a result, the IceCube data above 100 TeV energies can be explained by neutrinos produced via $pp$ interactions (see Fig.~2 left). The model predicts a neutrino spectral break around PeV energies, above which the the neutrino spectrum is steepened by the diffusion process. Gamma rays produced in cluster environments are cascaded down, leading to GeV-TeV gamma rays. In this model, direct gamma rays can be significantly suppressed because low-energy cosmic rays lose their energies via adiabatic losses during the expansion of radio bubbles. Also, the spectrum of cosmic rays confined in the cluster environments should be harder than that of cosmic rays accelerated at the sources. 
The highest-energy cosmic rays escape into intergalactic space and can contribute to the observed cosmic-ray flux. Importantly, the fate of protons and nuclei are different. The nuclei are confined in the cluster environments for a long time, and they can be photodisintegrated due to the intracluster infrared background. As a result, the spectrum of nuclei is harder than that of protons, which enables us to explain the UHECR data including the chemical composition (see Fig.~2 right) as well as the Galactic-extragalactic transition around the second knee at ${10}^{17}$~eV. 
Notably, it has been proposed that galaxy clusters with/without radio galaxies are responsible for cosmic rays above the second knee, and the ``predicted'' high-energy neutrino flux is consistent with the observed IceCube data above 100 TeV energies~\cite{Murase:2008yt,Kotera:2009ms}. As shown in Ref.~\cite{Fang:2017zjf}, grand-unification is achieved as a natural extension of this model.    
During intergalactic propagation, UHECRs should produce cosmogenic neutrinos and gamma rays via $p\gamma$ interactions with the cosmic microwave background and extragalactic background light. 
The grand-unification scenario predicts a smooth transition from the source neutrino flux to the cosmogenic neutrino flux with $\sim3\times{10}^{-9}~{\rm GeV}~{\rm cm}^{-2}~{\rm s}^{-1}~{\rm sr}^{-1}$ at EeV energies~\cite{Murase:2016gly,Fang:2017zjf}.  

One of the appealing points of the cosmic-ray reservoir models for IceCube neutrinos is their strong testability. 
Starburst galaxies are expected to have a high gas density, in which all cosmic rays are depleted via $pp$ interactions. Nearby starburst galaxies, M82 and NGC 253, are prototypical starbursts that may not allow sufficiently high-energy cosmic rays to be confined in the system. The calorimetric model predicts that more luminous starbursts like NGC 1068 and NGC 2146 are the most promising, and NGC1068 is the best candidate among nearby starbursts in the northern sky~\cite{Murase:2016gly}. The IceCube data clearly indicate that cosmic rays need to be accelerated up to $\sim100$~PeV energies that is beyond the knee at $3\times{10}^{15}$~eV~\cite{Murase:2013rfa}. This requires the existence of cosmic-rat accelerators that are more powerful than ordinary supernovae, and proposed possibilities include hypernovae, GRBs, jets and winds from AGNs~\cite{Murase:2013rfa,Tamborra:2014xia}. 
NGC 1068 is the most interesting source in this aspect, because it coexists with a Seyfert galaxy that may be accompanied by disk-driven winds~\cite{Liu:2017bjr}. The detectability of muon neutrinos in ten-year operations of IceCube is shown in Fig.~3 left. 
Similarly, nearby galaxy clusters are promising sources of neutrinos especially if the galaxy clusters/group model for IceCube neutrinos is correct (see Fig.~3 right). Theoretical predictions depend on the spatial distribution of cosmic rays inside clusters, and the Virgo cluster is expected to be among the most promising sources~\cite{Murase:2012rd}. 
In either of starburst or galaxy cluster model, predicted fluxes for individual nearby sources should be tested by IceCube-Gen2 through either clustering or stacking searches. These sources are transparent to GeV-TeV gamma rays, and observations with {\it Fermi} and various imaging atmospheric Cherenkov telescopes including CTA are critical to test the grand-unification scenario~\cite{Murase:2016gly}. 

\begin{figure}[tb!]
\begin{center}
\begin{tabular}{cc}
\includegraphics[height=5.0cm]{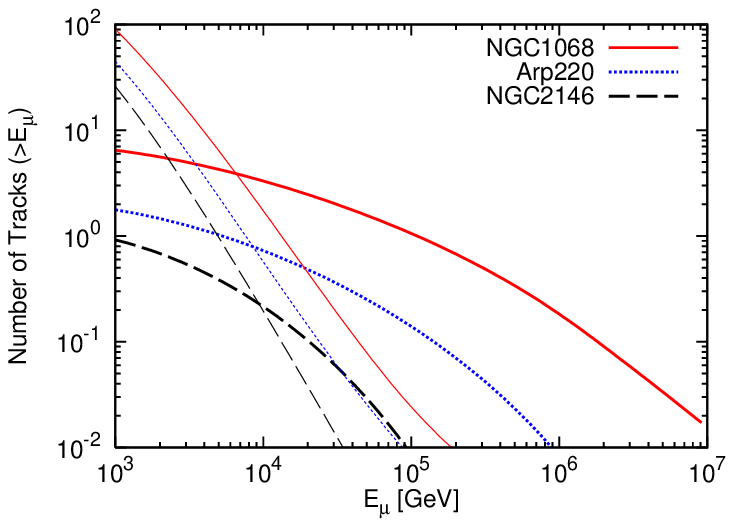} 
\includegraphics[height=5.0cm]{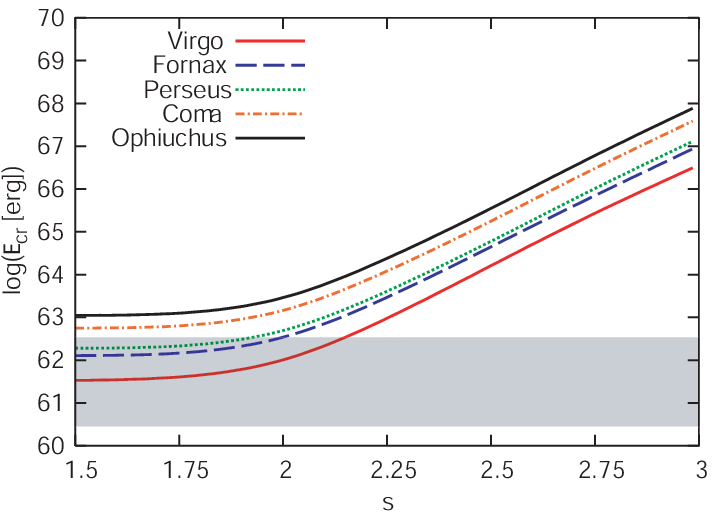} 
\end{tabular}
\end{center}
\vspace{-1.5em}
\caption{
\textsl{Left: The number of muon tracks expected for representative starburst galaxies, Arp 220, NGC 1068 and 2146~\cite{Murase:2016gly}. Ten years of operations by IceCube are assumed. Muon neutrino fluxes are normalized by the {\it Fermi} gamma-ray fluxes at GeV energies. The sum of the atmospheric and astrophysical neutrino backgrounds (dashed and dot-dashed) are also shown.\\
Right: Constraints on the total energy of cosmic rays confined in galaxy clusters for five nearby galaxy clusters~\cite{Murase:2012rd}. Five years of operations by IceCube are assumed. The grey shaded region represents the typical range of the cosmic-ray energy expected in cosmic-ray reservoir models that can explain the cosmic-ray flux above the second knee and the IceCube data above 100~TeV.}}
\vspace{-0.5em}
\label{fig:cosmicray}
\end{figure}

\section{Hidden Cosmic-Ray Accelerators and MeV Gamma-Ray Connection}
The latest observations suggest that the 10-100 TeV neutrino flux is higher than the flux above 100 TeV energies, which suggests the existence of hidden neutrino sources~\cite{Murase:2015xka}. 
First, this is because the predicted gamma-ray flux associated with the diffuse neutrino flux overshoots the diffuse isotropic gamma-ray background measured by {\it Fermi}. 
This is especially the case when neutrinos are produced via $pp$ interactions as expected in cosmic-ray reservoir models. 
Second, if the dominant neutrino production channel is the photomeson production process, there is a one-to-one correspondence between the optical depth to $\gamma\gamma$ interactions and that to $p\gamma$ interactions. 
The effective optical depth to $p\gamma$ interactions has to be larger than $\sim0.01$ not to overshoot the observed cosmic-ray flux and/or to satisfy energy budget requirements with known source classes. The ratio of the two cross sections imply that the optical depth to $\gamma\gamma$ interactions has to be larger than $\sim10$, which suggests that the sources are optically thick for GeV-TeV gamma rays. Thus, hidden cosmic-ray accelerators can naturally avoid the overshooting problem. 

GRBs are the brightest explosive phenomena in the Universe. 
However, stacking analyses on GRB prompt emission have shown that canonical GRBs can contribute to the observed diffuse neutrino flux only up to $\sim1$\%. So typical high-luminosity GRBs can only be subdominant in the diffuse neutrino sky. 
The story is different for population of dim GRBs such as low-luminosity GRBs. Indeed, some of the theoretical predictions for neutrinos from low-luminosity GRBs are consistent with the IceCube data above 100~TeV energies~\cite{Murase:2006mm}. 
However, the explanation for 10-100~TeV neutrino data observed in IceCube is more challenging. From the astrophysical view point, it is natural to expect that there is population of supernovae harboring choked jets. GRBs can occur only if a relativistic jet successfully penetrates a progenitor star, but it can readily be ``choked'', depending on the jet luminosity, opening angle, duration, and external medium. For example, if the central engine is not powerful or extended material exists, the jet can easily get stuck without breakout. If cosmic rays can still be accelerated by the shock acceleration mechanism or magnetic reconnections, the choked GRB jets should serve as efficient neutrino emitters without being accompanied by hadronic gamma rays~\cite{Meszaros:2001ms}. However, the cosmic-ray acceleration in radiation dominated environments should be considered with great caution. It is known that the diffusive shock acceleration is inefficient if the shock is radiation mediated. Because this necessary condition prohibits efficient neutrino production in high-density environments, lower-power choked GRB jets are expected to be more promising as the sources of high-energy neutrinos~\cite{Murase:2013ffa} (see Fig.~4 left). 
The models can be tested by stacking searches as have been done in Refs.~\cite{Senno:2017vtd,Esmaili:2018wnv} or neutrino-triggered follow-up observations as proposed in Ref.~\cite{Murase:2006mm}. 

The vicinity of AGNs, in which mass accretion flows onto a supermassive black hole exist, can be a promising site of efficient high-energy neutrino production (see Fig.~4 right). The recent simulations have suggested that particles could be accelerated in radiatively inefficient accretion flows or coronae via turbulence generated by the magnetorotational instability~\cite{Kimura:2016fjx,Kimura:2018clk}. The ion component in ion-electron plasma is expected to be collisionless in these environments, and protons and nuclei can be accelerated up to high energies via magnetic reconnections and stochastic acceleration. 
Efficient neutrino and gamma-ray production is unavoidable in such AGN core models, because strong radiation fields are known to exist. For example, for radio-quiet (RQ) AGNs (that are mostly Seyfert galaxies), X rays from their coronae provide target photons for the photomeson production. However, PeV cosmic rays responsible for 10-100~TeV neutrinos do not have sufficiently high energies for the photomeson production with disk photons in the ultraviolet range. Instead, the Bethe-Heitler process is dominant as energy losses of PeV cosmic rays, and  the total energy of gamma rays from proton-induced cascades is significantly larger than the total energy of high-energy neutrinos. 
Importantly, disk and X-ray spectra of RQ AGNs are well measured by optical and X-ray observations, which leads to a prediction of the unique multi-messenger connection between MeV gamma rays and medium-energy neutrinos (see Fig.~5 left). 
RQ AGNs may be responsible for the extragalactic X-ray and MeV gamma-ray backgrounds as well as the medium-energy neutrino flux, while RL AGNs including blazars may explain the extragalactic gamma-ray background from MeV to sub-TeV energies, and galaxy clusters with RL AGNs may explain the sub-PeV neutrino flux and the non-blazar component of the sub-TeV gamma-ray background.     

\begin{figure}[tb!]
\begin{center}
\begin{tabular}{cc}
\includegraphics[height=3.5cm]{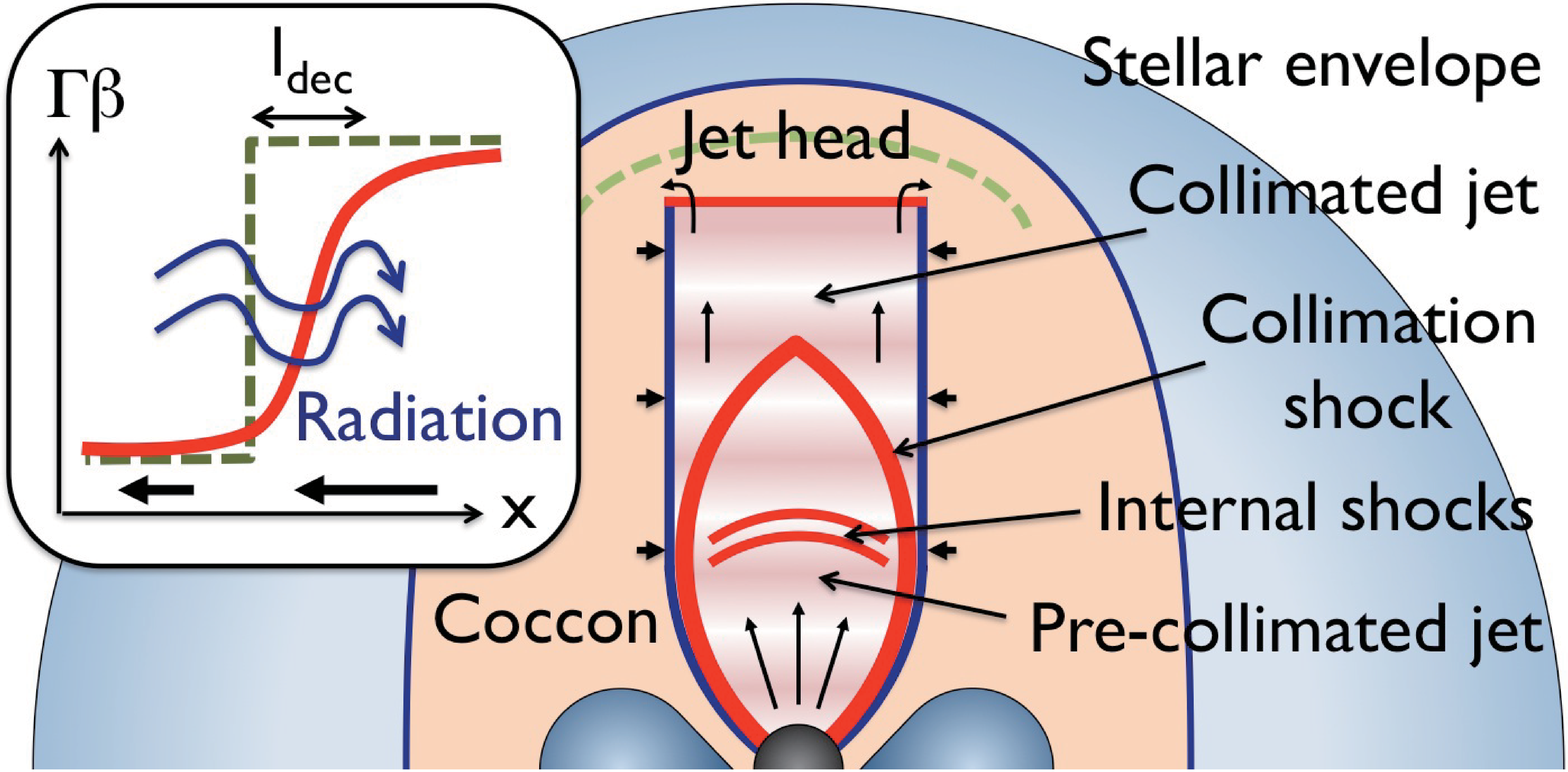} 
\includegraphics[height=3.8cm]{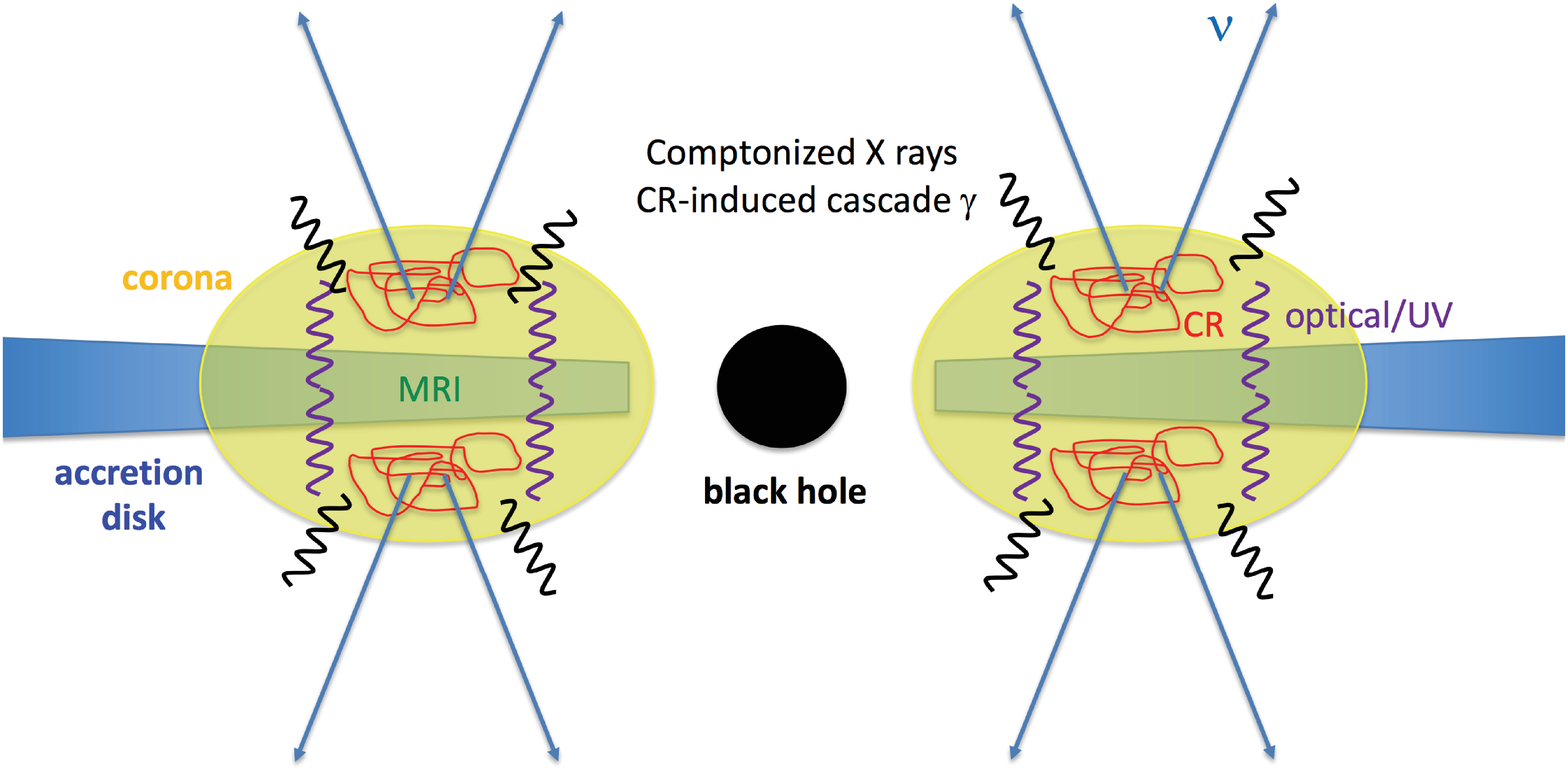} 
\end{tabular}
\end{center}
\vspace{-1.5em}
\caption{
\textsl{Left: Schematic picture of choked GRB jet models~\cite{Murase:2013ffa}. Cosmic rays can be accelerated in internal shocks or collimation shocks if shocks are not mediated by radiation. Sufficiently high-energy cosmic rays are all depleted for meson production. Only neutrinos can directly escape from the system.\\
Right: Schematic picture of AGN core models~\cite{Murase:2019vdl}. Cosmic rays can be accelerated in the coronal region above an accretion disk around a supermassive black hole. Neutrinos can escape while gamma rays are subject to electromagnetic cascades.}}
\vspace{-0.5em}
\label{fig:cosmicray}
\end{figure}

\begin{figure}[tb!]
\begin{center}
\begin{tabular}{cc}
\includegraphics[height=5.0cm]{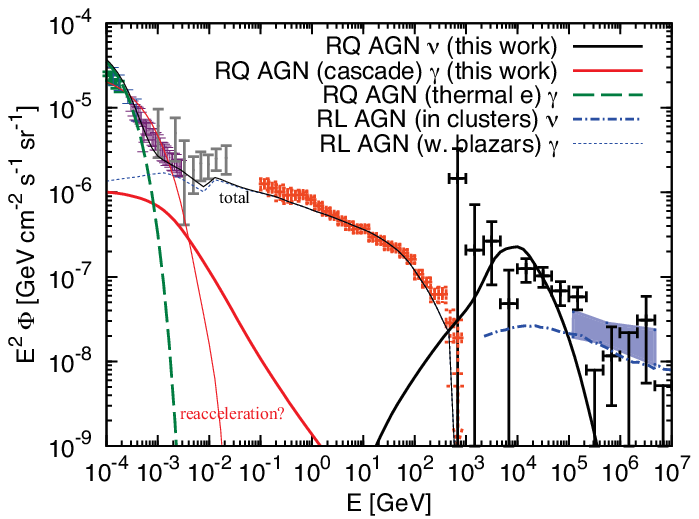} 
\includegraphics[height=5.0cm]{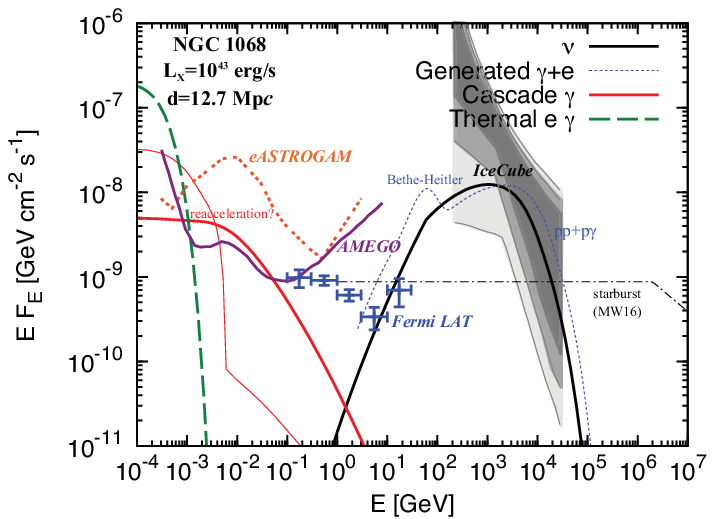} 
\end{tabular}
\end{center}
\vspace{-1.5em}
\caption{
\textsl{Left: Diffuse neutrino and gamma-ray fluxes expected in the AGN corona model, in which particles are accelerated in the coronal region of Seyfert galaxies~\cite{Murase:2019vdl}. High-energy coronal emissions are responsible for 10-100~TeV neutrinos and MeV gamma rays, as well as the X-ray background in this model. The higher-energy neutrino flux and GeV-TeV gamma rays originate from RL AGNs and their environments.\\
Right: Point source fluxes of neutrinos and gamma rays for a single RQ AGN such as NGC 1068~\cite{Murase:2019vdl}. The source distance is set to $d=12.7$~Mpc. The effect of reacceleration of secondary pairs, which could occur for certain parameters, is also shown assuming a reacceleration efficiency of $0.03$\%.}}
\vspace{-0.5em}
\label{fig:cosmicray}
\end{figure}

Radiation fields used for the photomeson and Bethe-Heitler pair production processes inevitably lead to proton-induced electromagnetic cascades. It is shown that the pairs are mainly radiated as MeV gamma rays, whether the dominant process is inverse-Compton radiation or synchrotron emission. The model predicts that RQ AGNs within dozens of Mpc should be detectable by future MeV gamma-ray detectors such as AMEGO and eASTROGAM (see Fig.~5 right). Associated high-energy neutrino emission can also be observed by IceCube-Gen2. 
Furthermore, not only individual point source searches but also stacking and clustering analyses should be powerful to test the AGN corona model for IceCube neutrinos.

\section{Summary}
We discussed the origin of IceCube neutrinos in light of multi-messenger connections. 

The diffuse neutrino flux above 100 TeV energies can naturally be explained by cosmic-ray reservoir models.  
In particular, starburst galaxies and galaxy clusters/groups provide examples of the grand-unification scenario, in which sub-PeV neutrinos, sub-TeV gamma rays, and UHECRs come from the same source population.
An important aspect of this scenario is testability. Both galaxy cluster/group and starburst models predict that the effective source number density is $n_s\sim{10}^{-5}~{\rm Mpc}^{-3}$~\cite{Murase:2016gly,Fang:2017zjf}, which can be examined by IceCube-Gen2 via clustering and stacking searches.  

The diffuse neutrino flux below 100 TeV energies is likely to originate from hidden cosmic-ray accelerators. Choked GRB jets associated with supernovae and AGN cores associated with Seyferts or low-luminosity AGNs seem promising candidates. The choked GRB jet models can be tested by upcoming stacking searches for nearby supernovae. AGN core models also have clear testable predictions - nearby Seyferts and even low-luminosity AGNs are detectable with IceCube-Gen2. These models also predict the unique multi-messenger connection between 10-100 TeV neutrinos and MeV gamma rays, which should be confirmed or falsified by future MeV gamma-ray telescopes such as AMEGO and eASTROGAM as well as IceCube-Gen2. 

\bibliographystyle{hunsrt}
\bibliography{kmurase.bib}

\end{document}